\documentclass[nofootinbib,groupedaddress,superscriptaddress,unsortedaddress,showpacs]{article}
\usepackage{amsmath,amsfonts,slashed}
\usepackage{epsfig,epsf}
\usepackage{hyperref}
\usepackage{float}
\restylefloat{table}

\usepackage[dvips]{color}
\usepackage[normalem]{ulem}
\definecolor{darkgreen}{cmyk}{1,0,1,0.4}

\definecolor{darkred}{cmyk}{0,1,1,0.4}

\usepackage{slashed}

\def\gsim{\buildrel{\scriptscriptstyle >}\over{\scriptscriptstyle\sim}}
\newcommand{\wt}{\widetilde}

\def \chonjpm{{\wt\chi_j}^{\pm}}
\def \chonepm{{\wt\chi_1}^{\pm}}

\def \mchonepm{m_{\chonepm}}

\def \chtwopm{{\wt\chi_2}^{\pm}}
\def \mchtwopm{m_{\chtwopm}}

\def \lspi{\wt\chi_i^0}

\def \lspone{\wt\chi_1^0}
\def \mlspone{m_{\lspone}}
\def \lsptwo{\wt\chi_2^0}

\def \lspthree{\wt\chi_3^0}

\def \lspfour{\wt\chi_4^0}
\def \mlspfour{m_{\lspfour}}

\def \slepl{\wt{l}_L}
\def \mslepl{m_{\slepl}}
\def \slepr{\wt{l}_R}
\def \mslepr{m_{\slepr}}
\def \met  {\mbox{${E\!\!\!\!/_T}$}}
\def\gmin2{(g-2)_\mu}
\begin{document}
\renewcommand*{\thefootnote}{\fnsymbol{footnote}}

\begin{center}
{\Large \bf The past, present and future of the heavier electroweakinos in the light of LHC and other data }\\
\vspace*{1cm} 
{\sf Amitava Datta$^{a,}$\footnote{email: adatta\_ju@yahoo.co.in}},
{{\sf Nabanita Ganguly$^{b,}$\footnote{email: nabanita.rimpi@gmail.com}}  
} \\  
\vspace{10pt}  
{\small {\em $^a$ INSA Senior Scientist \\
Department of Physics, University of Calcutta,  
92 Acharya Prafulla Chandra Road,\\
Kolkata 700009, India \\
$^b$ Department of Physics, University of Calcutta,  
92 Acharya Prafulla Chandra Road,\\
Kolkata 700009, India}}
 
\normalsize  
\end{center}


\vspace{5mm}
\begin{abstract}\vspace*{10pt}

The aim of this paper is to showcase the novel multilepton ($nl + \met$, n = 3 - 5) signals, hitherto unexplored at the LHC, arising from the heavier electroweakinos, in several generic pMSSMs at the upcoming LHC experiments. We first briefly review our old constraints on the full electroweakino sector of these models, containing both lighter and heavier sparticles, using the ATLAS trilepton data from the LHC Run I. Next we derive new stronger constraints on this sector for the first time using the ATLAS Run II data. We identify some benchmark points and explore the prospect of observing multilepton events in future LHC experiments. Our focus is on the channels with $n > 3$ which are the hallmarks of the heavier electroweakinos. If the spectrum of the lighter electroweakinos is compressed, these signals might very well be the discovery channels of the electroweakinos at the high luminosity LHC. We also discuss the implications of the new LHC constraints for the observed dark matter relic density of the universe, the measured value of the anomalous magnetic moment of the muon and the dark matter direct detection experiments.  

\vskip 5pt \noindent  
\texttt{PACS Nos:12.60.Jv, 14.80.Nb, 14.80.Ly, 95.35.+d, 13.85.-t
  } \\  
\end{abstract}

\renewcommand*{\thefootnote}{\arabic{footnote}}
\setcounter{footnote}{0}

\section{Introduction}
Supersymmetry (SUSY) is a novel symmetry which predicts that corresponding to every boson (fermion) in the Standard Model (SM) there
is a fermionic (bosonic) superpartner which are collectively called the sparticles (For reviews and text books on supersymmetry, see, {\it e.g.}, \cite{Nilles:1983ge, Haber:1984rc, Martin:1997ns, Chung:2003fi} and \cite{susybook1,susybook2} respectively). The painstaking searches for the sparticles spanning several years at the LHC Run I and Run II experiments are approaching the next long shutdown. Yet no signal has been seen so far. This leads to stringent lower bounds on many sparticle masses \cite{atlastwiki,cmstwiki}. As expected the bounds on the masses of the strongly interacting sparticles (the squarks and the gluinos)  with large production cross-sections turn out to be the most stringent ones. In some models the relevant limits could be as large as 2 - 3 TeV. Therefore the possibility that the masses of these sparticles could be beyond the kinematic reach of the LHC is gradually gaining  ground. 

If this indeed is the case then the best bet for SUSY discovery is to search for the spin-1/2 sparticles belonging to the electroweak (EW) sector. These  superpartners of the gauge and Higgs bosons are called  the electroweakinos (eweakinos). As a sequel to our earlier works \cite{Datta:2016ypd,Chakraborti:2017vxz}, we wish to highlight in this paper the novel multilepton ($nl + \met$, n = 3,4,5) signals arising from the heavier ones among the eweakinos  in several  generic models at the upcoming LHC experiments. It may be stressed that the signals for $ n>3$ are hallmarks of the heavier eweakinos as the strength of these signals are rather poor if they are decoupled. Moreover, if the lighter eweakinos have a compressed spectrum these could very well be eweakino discovery channels.

It is worth recalling that the LHC collaborations have so far executed dedicated searches in the trilepton channel targeting the lighter eweakinos using both Run I \cite{Aad:2014nua, Aad:2014vma, Khachatryan:2014qwa, Khachatryan:2014mma} and Run II \cite{Sirunyan:2017lae, Aaboud:2018jiw} data. As is well known model independent mass limits are hard to extract from the current data since the signals depend on too many unknown parameters (mostly the soft SUSY breaking terms) present in the most general Minimal Supersymmetric Standard Model (MSSM). Thus the LHC collaborations usually derive the constraints from the search results in the so called simplified models \cite{Aad:2014nua, Aad:2014vma, Khachatryan:2014qwa, Khachatryan:2014mma, Sirunyan:2017lae, Aaboud:2018jiw}. These models may be obtained after imposing some simplifying assumptions on the general  MSSM which reduce the number of free parameters. Decoupling of the heavier eweakinos is one of the many ad hoc assumptions thus invoked.  

The above limits were reexamined \cite{Chakraborti:2014gea, Chakraborti:2015mra} in the phenomenological MSSM (pMSSM) \cite{Djouadi:1998di} with 19 free parameters. It has been shown that in some regions of the parameter space the predictions of the pMSSM resemble that of the simplified models employed by the ATLAS group quite well and the resulting limits are very similar (for comparisons using Run I and Run II data, see Fig. 1 of \cite{Chakraborti:2014gea}, Fig. 7,8 of \cite{Choudhury:2016lku} and Fig. 1 of this paper). In several other regions, however, the limits in the pMSSMs are significantly weaker. However, the decoupling of heavier eweakinos was also assumed in these papers. In fact most of the recent analyses involving the eweakinos \cite{Choudhury:2013jpa, Eckel:2014dza, Han:2014xoa, Han:2014sya, Barman:2016kgt} also imposed the ad hoc assumption that the heavier eweakinos  are decoupled. 

The heavier eweakinos were included in the analyses of \cite{Datta:2016ypd, Chakraborti:2017vxz, Martin:2014qra, Arina:2016rbb} using the LHC Run I data and very recently in \cite{Athron:2018vxy} using the LHC Run II data. In the detailed analyses of \cite{Datta:2016ypd, Chakraborti:2017vxz} it was pointed out that the non-decoupled heavier eweakinos may have three important implications for the LHC searches.
\begin{itemize}
\item The ATLAS and CMS collaborations have interpreted the null search results from the $3l + \met$ signal in various simplified models with decoupled heavier eweakinos. The main results of their  analyses are exclusion contours in the $\mchonepm - \mlspone$ plane. On the other hand in a pMSSM with non-decoupled heavier eweakinos similar constraints may become significantly stronger due to the additional contributions from the heavier eweakinos to the signal (see Figs. 3,4 and 5 of \cite{Chakraborti:2017vxz} based on ATLAS Run I data). This, however, is a quantitative change. 
\item There are qualitatively new results as well. The cascade decays of the heavier eweakinos can lead to novel multilepton (n-lepton $(l) + \met$, n = 3,4,5) signals. It may be recalled that events with $n > 3$ are not very common in the models with decoupled heavier eweakinos.
\item If the lighter eweakinos have a compressed mass spectrum the signals  from the heavier one could even be the SUSY discovery channels. For example, the conventional trilepton signals ($n=3$) which dominantly come from the former may be swamped by the SM background while signals with
$n > 3$ triggered by the latter, which have highly suppressed backgrounds, may show up at the LHC. 
\end{itemize}
The last two points were illustrated in section 6 of \cite{Chakraborti:2017vxz}.

It should be emphasized that the interest in the eweakino  sector is not restricted to LHC signatures alone. 
These sparticles can shed light on the origin of the observed Dark Matter (DM) in the Universe \cite{Hinshaw, Ade:2015xua}\footnote{For reviews and recent phenomenological works see {\it e.g.,} \cite{Baer:2008uu, Calibbi:2013poa, Demirci:2014gva, Calibbi:2014lga, Roszkowski:2014iqa, Bramante:2014tba, Choudhury:2015lha, Bramante:2015una, Hamaguchi:2015rxa, Cao:2015efs, Chattopadhyay:2016ivr, Chakraborti:2017dpu, Abdughani:2017dqs}}, improve the agreement between the measured anomalous magnetic moment of the muon ($a_{\mu}$) \cite{Bennett:2006fi, Roberts:2010cj} and the theoretical prediction \cite{Jegerlehner:2009ry, Hagiwara:2011af}. Last but not the least, the naturalness \cite{Sakai:1981gr, Kaul:1981hi, Barbieri:1987fn, Feng:2013pwa} of any SUSY model favours small values of the EW parameter $\mu$ known as the higgsino mass parameter. The constraints on this parameter from the LHC searches and other observables can , therefore, potentially test various SUSY models in the light of naturalness arguments.

In this paper we update and upgrade the constraints in \cite{Datta:2016ypd, Chakraborti:2017vxz} using, for the first time, the LHC Run II data (ATLAS) and other non LHC constraints, taking into account all eweakinos - the heavier as well as the lighter ones. We then define a set of post LHC Run II  benchmark points (BPs) and use them to assess the prospect of observing the multilepton signatures in future high luminosity LHC experiments after the next long shut down.

The plan of this paper is as follows. In section \ref{pmssmmodels} we present a brief discussion of different pMSSMs involving both heavier and lighter eweakinos studied in this work. The models are summarized in Table \ref{summarytab} and the choice of parameters for scanning in each case is listed after this table. The methodology adopted to get the main results are described in detail in section \ref{methodology}. In section \ref{result} we identify the allowed parameter space (APS) of the models discussed in section \ref{pmssmmodels} in the light of LHC data from Run II, the observed value of DM relic density of the universe and also the experimental constraint from the measured value of the anomalous magnetic moment of muon. The prospect of observing various multilepton signals in different models is assessed using post LHC Run II BPs selected from the corresponding APS. In section \ref{drctdtctn} we check the status of all models introduced in section \ref{pmssmmodels} vis-a-vis the spin-independent DM direct detection cross-section limits. Finally we conclude in section \ref{conclusion}.  

\section{The pMSSMs to be constrained}
\label{pmssmmodels}
In this section we briefly review several pMSSMs with 19 parameters \cite{Djouadi:1998di} which are then  constrained using the LHC eweakino search at Run II and other data in a later section. We emphasize that these models are generic in the sense that different models are characterized by certain hierarchies among the masses and mass parameters rather than their specific values. The fermionic sparticles in the EW sector are the charginos ($\chonjpm$, $j= 1, 2$) and the neutralinos ($\lspi$, $i = 1 - 4$) - collectively called the eweakinos. The indices i and j are arranged in ascending order of masses. The masses and the compositions of these sparticles are determined by four parameters: the U(1) gaugino mass parameter $M_1$, the SU(2) gaugino mass parameter $M_2$, the higgsino mass parameter $\mu$  and tan $\beta$ - the ratio of the vacuum expectation values of the two neutral Higgs bosons. If no assumption regarding the SUSY breaking mechanism is invoked, the soft breaking masses $M_1$, $M_2$ and the superpotential parameter $\mu$ are all independent. Throughout this paper we take tan $\beta$ = 30 since relatively large values of this parameter give a better agreement with the $a_{\mu}$ data, ensure that the SM like Higgs boson has practically the maximum mass at the tree level and improve the prospect of charged Higgs boson search. The stable, neutral lightest neutralino ($\lspone$), which is assumed to be the lightest supersymmetric particle (LSP), is a popular DM candidate. 

The scalar sparticles are the $L$ and $R$ type sleptons which are superpartners of leptons with left and right chirality. The sneutrinos are the superpartners of the neutrinos. We assume L(R)-type sleptons of all flavours to be  mass degenerate with a common mass $\mslepl$($\mslepr$). Because of the SU(2) symmetry the sneutrinos are mass  degenerate with L-sleptons modulo the D-term contribution. We neglect L-R mixing in the slepton sector. For simplicity we work in the decoupling regime (See {\it e.g.}, \cite{Djouadi:2005gj}) of the Higgs sector with only one light SM like Higgs boson, a scenario consistent with all Higgs data collected so far (See {\it e.g.}, \cite{Bechtle:2016kui}).

The signals of the eweakinos at the LHC are also sensitive to their compositions which are governed by the hierarchy among the parameters $M_1, M_2$ and $\mu$. Most of the existing analyses revolve around the broad scenarios listed in the next few subsections.

Following our earlier works \cite{Datta:2016ypd, Chakraborti:2017vxz, Chakraborti:2014gea, Chakraborti:2015mra} we introduce a convenient nomenclature with four letters for denoting the pMSSMs belonging to three broad scenarios. The first two letters  represent the composition of the lighter eweakinos which lead to the signals when the heavier ones are decoupled.  We have considered three generic cases: the LW (Light Wino) model ($M_2 << \mu$), the LH (Light Higgsino) model ($M_2 >> \mu$) and the LM (Light Mixed) model ($M_2 \approx \mu$). These models will be described in subsections \ref{winomodel}, \ref{hgsnmodel} and \ref{mixedmodel} respectively. In subsection \ref{compmodel} we shall consider a few LH models where the lighter eweakino spectrum is compressed in different ways and the observable signals are mainly due to the heavier eweakinos.  
 
\subsection{The LW models ($M_2 << \mu$)}
\label{winomodel}
In this class of models the two relatively light and nearly degenerate eweakinos ($\chonepm$ and $\lsptwo$) are wino like and their masses are controlled by the parameter $M_2$. They are the main sources of the signal/signals. The production cross-section of the higgsino like heavier eweakinos ($\chtwopm$, $\lspthree$ and $\lspfour$), with masses controlled by the parameter $\mu$, are suppressed both due to their composition and larger masses. Thus their contributions to the signal are negligible. The assumption that the heavier eweakinos are decoupled is therefore realistic in this case. Here the LSP is either a pure bino ($M_1 << M_2$) or a wino-bino admixture ($M_1 \approx  M_2$). The trilepton signal ($3l + \met$) in this model also depend sensitively on the hierarchy among the sleptons and the eweakino masses. This leads to the following subclasses: 
\begin{itemize}
\item LWLS (Light Wino Light Left Slepton) model (1.1 a).  
\item LWHS (Light Wino Heavy Slepton) model (1.1 b). 
\end{itemize}
The simplified model considered by the LHC collaborations \cite{Aad:2014nua, Aad:2014vma, Khachatryan:2014qwa, Khachatryan:2014mma, Sirunyan:2017lae, Aaboud:2018jiw} with wino dominated $\chonepm$ and $\lsptwo$, bino dominated $\lspone$ and decoupled heavier eweakinos is a special case of this generic pMSSM in the limit of very large $\mu$. 

In the LWLS model (1.1 a) only the left sleptons ($\slepl$) are lighter than $\chonepm$ and $\lsptwo$ while the right sleptons ($\slepr$) are assumed to be decoupled. These eweakinos directly decay into sleptons and sneutrinos via two body modes with large BRs which enhances the leptonic signals. Sleptons belonging to all generations are assumed to be degenerate and their common mass lies between $\mlspone$ and $\mchonepm$. The choice $m_{\tilde l_L} = (\mchonepm+\mlspone)/2$ by the LHC collaborations optimizes the leptonic signals and yields the strongest bounds on the lighter eweakino masses (see section \ref{simulation}, Fig. 1). In later section we shall mostly use this choice of $m_{\tilde{l}_L}$ whenever this sparticle is assumed to be light. However, one can also think of various tilted scenarios where $\mslepl$ is either shifted towards $\mlspone$ or $\mchonepm$ so that the eweakino spectrum is somewhat compressed leading to weaker but not drastically different mass limits if the compression is not extreme. Several tilted models were examined in the light of LHC Run I data and other constraints \cite{Chakraborti:2014gea}.

In the LWHS (1.1 b) model all sleptons ($\slepl$ and $\slepr$) are heavier than $\chonepm$ and $\lsptwo$. These eweakinos decay into leptonic final states only via (on-shell or off-shell) $W$ and $Z$ bosons respectively. Since the branching ratios (BRs) of leptonic $W$ and $Z$ decays are small, the  leptonic signals in this case are suppressed compared to the LWLS model leading to weaker bounds on $\mchonepm$. The LHC collaborations have published mass limits in a simplified model related to this scenario assuming decoupled heavier eweakinos \cite{Sirunyan:2017lae, Aaboud:2018jiw}. Multilepton signals are not favoured in these LW type models. However we will briefly discuss in a later section that these are one of those few models which are consistent with the current DM direct detection data \cite{Akerib:2016vxi, Aprile:2017iyp, Cui:2017nnn} taken at its face value.

\subsection{The LH models ($M_2 >> \mu$)}
\label{hgsnmodel}
In this class of models the relatively light higgsino like eweakinos are $\chonepm$, $\lsptwo$ and $\lspthree$ with masses  controlled by the parameter $\mu$. They are the main sources of the signal/signals if the heavier eweakinos are decoupled. The pair production cross-section of these higgsino like eweakinos  are small compared to that in the LW models for comparable masses of the lighter eweakinos. Thus weaker mass bounds are obtained from the LHC data. In all cases  the LSP is either a pure bino ($M_1 << \mu$) or a bino-higgsino admixture ($M_1 \approx \mu$). The constraints on this model using the Run I data were obtained in \cite{Chakraborti:2015mra}.

It should be stressed that the wino dominated heavier eweakinos ($\chtwopm$ and $\lspfour$) are phenomenologically important in this scenario. Their masses are determined by the free parameter $M_2$. As expected the pair production cross-section of these eweakinos are suppressed due to their larger masses. However, their favourable couplings to the gauge bosons compensate this suppression to some extent. As a result their contributions to the signals turn out to be appreciable or even dominant when the lighter eweakino spectrum is compressed. This point was emphasized in \cite{Datta:2016ypd, Chakraborti:2017vxz} and the importance of the heavier eweakinos was illustrated using the LHC RUN I data. We have constrained the following models using the Run II and other data in this paper : 
\begin{itemize}
\item The LHLS (Light Higgsino Light Left Slepton) model (2.2 a). 
\item The LHHS (Light Higgsino Heavy Slepton) model (2.2 b).
\end{itemize}
In the former model the L-slepton - lighter eweakino mass hierarchies are similar to that in the LWLS model (see subsection \ref{winomodel}). In the LHHS model it is  assumed that all sleptons are heavier than the lighter eweakinos ($\chonepm$, $\lsptwo$ and  $\lspthree$) but are lighter than the heavier eweakinos. In the numerical computations the common slepton mass is chosen to be $m_{\tilde l_L} = m_{\tilde l_R} = (\mchonepm + \mchtwopm)/2$ and we set $M_2 = 1.5 \mu$. An additional attraction of the LHHS model is that it is consistent with the DM direct detection data \cite{Akerib:2016vxi, Aprile:2017iyp, Cui:2017nnn} as will be shown in a later section.

\subsection{The LM models ($M_2 \approx \mu$)}
\label{mixedmodel}
Here all eweakinos except for the LSP are wino-higgsino admixtures. The LSP is dominantly a pure bino but in some zones of the parameter space all eweakinos are admixtures of all the weak eigenstates. In \cite{Chakraborti:2017vxz} the LMLS model was constrained using the LHC Run I data. In this paper we have updated these constraints using the LHC Run II data. 

\subsection{The Compressed LHHS models}
\label{compmodel}
In this section we consider a few LHHS models (2.2 b) where the lighter eweakinos have a compressed spectrum. As a result observable multilepton  signals come mainly from the heavier eweakinos. We consider the following models:  
\begin{itemize}
\item The CLHHS ($\wt{W}$) (Compressed Light Higgsino Heavy Slepton) model with wino ($\wt{W}$) like heavier eweakinos \cite{Datta:2016ypd, Chakraborti:2017vxz} (2.4 a).
\item The MCLHHS ($\wt{W}$) : Same as (2.4 a) except that the light higgsinos are moderately compressed \cite{Chakraborti:2017vxz} (2.4 b).  
\item The CLHHS ({$\wt{B} - \wt{W}$}) model with one bino ($\tilde{B}$) like and one wino ($\tilde{W}$) like heavier eweakino (2.4 c).
\end{itemize}

In the CLHHS ($\wt{W}$) model we set $M_1\simeq \mu$ with $M_2 >\mu$. This choice  leads to a compressed lighter eweakino spectrum where $\lspone$, $\lsptwo$, $\lspthree$ and $\chonepm$ are approximately mass degenerate and each has significant bino and higgsino components. The masses of the wino dominated heavier eweakinos are determined by the free parameter $M_2$. As in all LHHS models we set $m_{\tilde l_L} = m_{\tilde l_R} = (\mchonepm+\mchtwopm)/2 $ so that the sleptons are always heavier than lighter eweakinos. For future use we define a compression parameter $x = \mu/M_1$ which represents the degree of compression. For numerical results in the CLHHS ($\wt{W}$) model we have chosen $x =$ 1.05. 

As discussed in detail in sections 5.1 and 6.5 of \cite{Chakraborti:2017vxz}, the compatibility of CLHHS ($\wt{W}$) model with the observed DM relic density is obtained for $\mchtwopm > 600$ GeV. On the other hand in the MCLHHS ($\wt{W}$) model with slightly larger value of $x$ (= 1.3) this compatibility is obtained for lower values of $\mchtwopm$ which ensure better signals. In \cite{Chakraborti:2017vxz} this issue was illustrated with some BPs. Here we make a detailed study of the phenomenology of this model by making a parameter space scan using the constraints from the LHC Run II and other data.
 
The CLHHS ($\wt{B}-\wt{W}$) model with non-decoupled heavier eweakinos have higgsino like and nearly degenerate $\lspone$, $\lsptwo$, $\lspthree$ and $\chonepm$. As a result the signals from the lighter eweakinos are expected to consist of only soft visible particles.  For a long time there was no LHC constraint on this model. More recently both ATLAS \cite{Aaboud:2017leg} and CMS \cite{Sirunyan:2018iwl} collaborations have obtained some interesting constraints on simplified models closely related to this model using improved techniques for detecting soft leptons \cite{Schwaller:2013baa}. The excluded parameter space corresponds to $\mlspone \approx \mchonepm =$ 100 - 140 GeV. A comparison with Fig. 5 of \cite{Chakraborti:2015mra} shows that in a closely related pMSSM such masses may be theoretically forbidden. In this paper we focus on scenarios with non-decoupled heavier eweakinos. Here $\chtwopm$($\lspfour$), $\lspthree$ are chosen to be wino and bino dominated respectively or admixtures of these components. Then multilepton signals can directly come from the production and decay of these sparticles. 

\subsection{Summary of  parameter spaces in different pMSSMs and the method of scanning}
\label{paramsummary}
We summarize the mass parameter hierarchy and the corresponding compositions of eweakinos in Table \ref{summarytab} for each pMSSM discussed in this section: 

\begin{table}[H]
\begin{center}
\begin{tabular}{|c|c|c|c|c|}
\hline
\hline
Model & Hierarchies among &\multicolumn{3}{c|}{Compositions of eweakinos}\\
\cline{3-5}
Name &mass parameters &LSP &Lighter eweakinos &Heavier eweakinos \\
\cline{1-5}
LHLS &$M_1  < M_{\wt{l}} < \mu < M_2$ &Bino &Higgsino &Wino \\
(see model (2.2 a)) & & & & \\
\hline
LHHS &$M_1 < \mu < M_{\wt{l}} < M_2$ &Bino &Higgsino &Wino \\
(see model (2.2 b)) & & & & \\
\hline
LMLS &$M_1 < M_{\wt{l}} < M_2 \approx \mu$ &Higgsino &Wino-higgsino &Wino-higgsino \\
(see subsection \ref{mixedmodel}) & & & & \\
\hline
CLHHS ($\wt{W}$), MCLHHS ($\wt{W}$)  &$M_1 \approx \mu < M_{\wt{l}} < M_2$ &Bino-higgsino &Bino-higgsino &Wino \\
(see models (2.4 a) and (2.4 b)) & & & &\\
\hline
CLHHS ($\wt{B}-\wt{W}$)  &$\mu < M_{\wt{l}} < M_1 = M_2$ &Higgsino &Higgsino &Bino, Wino \\
(see model (2.4 c)) & & & &\\
\hline
\hline
\end{tabular}
\end{center}
\caption{Hierarchies among mass parameters for different models described in detail in section \ref{pmssmmodels}. The compositions of eweakinos in each case are also shown.}
\label{summarytab}
\end{table} 
In order to carry out the parameter space scanning in each pMSSM to obtain the LHC limits, the following choices have been made for free and fixed parameters:
\begin{itemize}
\item {\bf {LHLS Model :}} In this case the scanning is done over $M_1$ and $\mu$ while $M_2$ and $M_{\wt{l}}$ are fixed by $M_2 = 1.5$ $\mu$ and $M_{\wt{l}} = (\mlspone + \mchonepm)/2$ respectively.
\item {\bf {LHHS Model :}} The choice of parameters for LHHS model is exactly same as in LHLS model. The only difference is in the choice of slepton mass parameter which is taken as $M_{\wt{l}} = (\mchonepm + \mchtwopm)/2$ in this case.
\item {\bf{LMLS model :}} In this case $M_1$ and $M_2$ are taken as free parameters. $M_2$ and $\mu$ are very closely spaced and are related by the choice $\mu = 1.05$ $M_2$. Sleptons lie between the LSP and the lighter eweakinos with the specific choice $M_{\wt{l}} = (\mlspone + \mchonepm)/2$.
\item {\bf {CLHHS ($\wt{W}$) and MCLHHS ($\wt{W}$) models :}} Here we take $M_1$ and $\mu$ to be related by $\mu = x M_1$ where $x$ is taken as 1.05 for extreme compression (CLHHS ($\wt{W}$) and 1.3 for moderate compression (MCLHHS ($\wt{W}$). The scanning is performed over $M_1$ and $M_2$ while the slepton mass parameter $M_{\wt{l}}$ is taken as the arithmetic mean of $\mchonepm$ and $\mchtwopm$.
\item {\bf{CLHHS ($\wt{B}-\wt{W}$) model :}} For this type of model, $\mu$ and $M_1$ (which is degenerate with $M_2$) are free parameters. $M_{\wt{l}}$ is a dependent parameter which is related to eweakino masses via the relation $M_{\wt{l}} = (\mchonepm + \mchtwopm)/2$.
\end{itemize}

\section{Methodology}
\label{methodology}
The  work in this paper is based on the following methodology.

\subsection{The constraints}
\label{constraints}
We first constrain the pMMSMs discussed in the previous section using the model independent ATLAS Run II data in the $3l + \met$ channel collected with 36.1 fb$^{-1}$ of integrated luminosity \cite{Aaboud:2018jiw}. We have also used the ATLAS Run II constraints from slepton search data \cite{Aaboud:2018jiw} when the model under consideration contains a light slepton. The constraint on the CLHHS ($\wt{B}-\wt{W}$)\footnote{See model (2.4 c) of subsection \ref{compmodel}.} also takes into account the ATLAS higgsino search data using the soft lepton detection technique \cite{Aaboud:2017leg}. However we have not simulated the last two signals. Instead we have simply rejected the points lying within the ATLAS exclusion contours.     

We have also used the WMAP/Planck constraints \cite{Hinshaw, Ade:2015xua} and that from the measured value of the anomalous magnetic moment of the muon \cite{Bennett:2006fi, Roberts:2010cj} following the discussions of \cite{Chakraborti:2017vxz}. We believe that the theoretical and experimental uncertainties in the above three constraints are relatively small. To clarify this statement further we note that the constraints from flavour physics can be applied to the MSSM only after imposing yet another assumption known as the minimal flavour violation. In a nutshell this implies that the mixing angles in the squark and quark sectors are the same. For a discussion on non-minimal flavour violation , see for example, ref \cite{Jager:2008fc}.

We have also taken into consideration the constraints from different experiments on direct detection of the DM \cite{Akerib:2016vxi, Aprile:2017iyp, Cui:2017nnn}. As is well known this data disfavours many SUSY models. However there are many assumptions, both theoretical and experimental, in the derivation of the spin-independent LSP-nucleon scattering cross-section $\sigma_{SI}$ (for a brief discussion see, e.g, section 4 of \cite{Chakraborti:2017vxz} and the references there in). Relaxing these assumptions may significantly lower the computed value of $\sigma_{SI}$. This makes the comparison of the theoretical prediction and the experimental upper bound on $\sigma_{SI}$ somewhat ambiguous. We have, therefore, not displayed the impact of these constraints in our main figures. They are discussed in a separate section. 
 
\subsection{The Simulation}
\label{simulation}
Using \verb+PYTHIA+ we simulate the $3l +\met$ events in the pMSSMs studied by us. We closely follow the ATLAS group for selection and isolation of signal objects \cite{Aaboud:2018jiw}. Jets are reconstructed using the anti-$k_{T}$ \cite{Cacciari:2008gp} algorithm with radius parameter R = 0.4 and they have $p_{T} > 20$ GeV, $|\eta| < 2.8$. Signal $e$ and $\mu$ are required to have $p_{T} > 10$ GeV and $|\eta| < $ 2.47 (2.5) for $e$ $(\mu)$. ATLAS has defined 11 signal regions (SRs) each characterized by a set of cuts. Some of these regions target slepton mediated decays of $\chonepm$ and $\lsptwo$ while others target $W$ and $Z$ mediated decays. The results are presented in terms of number of observed events in the $3l + \met$ channel in each SR and the corresponding number of SM backgrounds (see Table 13 and 14 of \cite{Aaboud:2018jiw}) extracted from the data. With these two numbers one can obtain the model independent upper bound on $N_{BSM}$ for each SR \cite{Choudhury:2017acn}. We have used this information to constrain the pMSSMs discussed in the last section.
\begin{figure}[h]
\centering
\includegraphics[width=0.6\textwidth]{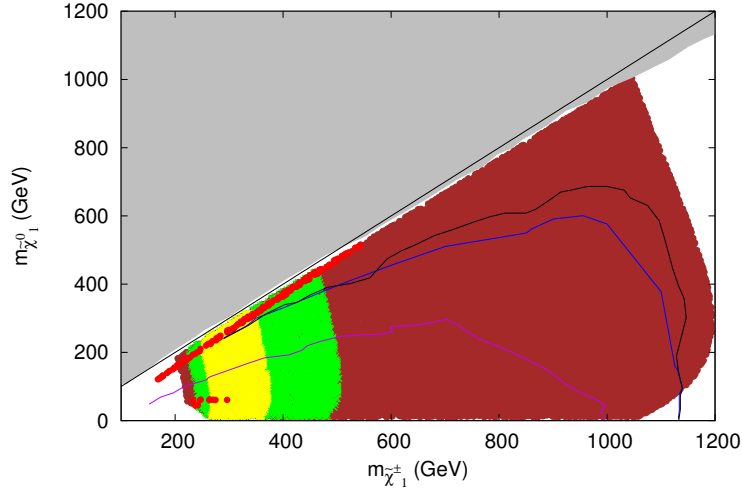}
\caption{The black line represents the exclusion contour in the $\mchonepm - \mlspone$ plane at 95$\%$ CL in a simplified model (see text) obtained by the ATLAS collaboration from trilepton searches at 13 TeV LHC \cite{Aaboud:2018jiw}. The blue line shows the exclusion obtained by our simulations in the closely related LWLS model. The area enclosed by the magenta curve is excluded by the ATLAS slepton search at Run II (see text). The brown, green and yellow regions are consistent with the $a_{\mu}$ data at $3\sigma$, $2\sigma$ and $1\sigma$ levels respectively. The red points satisfy WMAP/PLANCK data of the DM relic density. The grey region at the upper left corner is disfavoured theoretically.}
\label{fig}
\end{figure}  
In Fig. \ref{fig} we compare the exclusion contours obtained by us and the one by the ATLAS collaboration from the Run II trilepton search data \cite{Aaboud:2018jiw}. They had obtained the contour for a simplified model with wino like $\chonepm$ and $\lsptwo$ and a bino like LSP with light sleptons (the black exclusion contour). The pMSSM closest to the above simplified model is the LWLS model (see subsection \ref{winomodel}, model (1.1 a)) with decoupled heavier eweakinos. The blue exclusion contour in Fig. \ref{fig} is the result of our simulation in this model. It may be noted that for $\mchonepm >> \mlspone$, $\chonepm$ ($\lspone$) is almost a pure wino (bino) and the results of these two simulations agree quite well. As $\mlspone$ increases $\chonepm$ ($\lsptwo$) acquires significant bino component. As a result the $\chonepm$ $\lsptwo$ production cross-section decreases leading to weaker exclusions. Each point in the parameter space corresponds to a L-slepton mass due to the choice $m_{\tilde{l_L}} = (\mlspone + \mchonepm)/2$. The magenta curve is the exclusion contour from the ATLAS slepton search data at Run II.

The impact of the other two constraints - namely the measurements of $a_{\mu}$ and the DM relic density as discussed in section \ref{constraints} are also shown by different colour bands. The colour convention is explained in the figure caption. It may be noted that the APS consistent with all constraints is rather tiny. As we shall show in the next section the APSs in some of the LH models are considerably larger.

We next turn our attention to the prospect of observing multilepton signals ($nl + \met$ with n = 3,4,5) for an integrated luminosity of 3000 fb$^{-1}$. From the APS of each pMSSM  with non-decoupled heavier eweakinos we select a few BPs. We then simulate the signals corresponding to each BP for different n. We closely follow different selection criteria introduced in the ATLAS Run II analysis \cite{Aaboud:2018jiw}. The estimation of the SM background for each n is common to all pMSSMs studied here. They will be presented in the next section.

All signals in this work are generated using \verb+PYTHIA+ \cite{Sjostrand:2006za}. The relevant background processes in case of $nl + \met$ with $n > 3$ are generated using \verb+ALPGEN+ \cite{Mangano:2002ea} with MLM matching \cite{Hoche:2006ph, Mangano:2006rw} and then passed through \verb+PYTHIA+ for showering and hadronization. Jets are reconstructed using \verb+FASTJET+ \cite{Cacciari:2011ma} with anti-$k_{T}$ algorithm. For parton distribution function (PDF), CTEQ6L \cite{Pumplin:2002vw} has been used in all our simulations. 

\subsection{Scanning of the Parameter Spaces}
\label{scanning}
The squark mass parameters, $M_{A}$, $M_3$ which do not play any role in our present simulation are set at a large value of 2 TeV. The trilinear coupling $A_t$ is fixed at - 2 TeV so that the Higgs mass $m_h$ falls within the experimentally allowed window 122 GeV $< m_h <$ 128 GeV around a central value of 125 GeV \cite{Aad:2012tfa,Chatrchyan:2012xdj}. All other trilinear couplings are set at zero. The heavier Higgs like bosons are assumed to be decoupled. For all our simulations, we have fixed $\tan\beta$ at 30 which gives better agreement with $a_{\mu}$ data and the parameters $M_1$, $M_2$, $\mu$ are varied (for the details see subsection \ref{paramsummary} and Table \ref{summarytab}). The masses of sleptons are fixed by the definition of each model as discussed in section \ref{pmssmmodels}. The SM Parameters are taken as follows : $m_t^{pole} = 175$ GeV, $m_Z = 91.18$ GeV, $m_b^{\overline{ms}} = 4.2$ GeV and $m_{\tau} = 1.77$ GeV. The complete SUSY spectrum and $a_{\mu}$ are evaluated using \verb+SuSpect+ \cite{Djouadi:2002ze}. The decay modes of sparticles are calculated using \verb+SUSY-HIT+ \cite{Djouadi:2006bz}. We compute DM relic density and $\sigma_{SI}$ using \verb+micrOMEGAs+ \cite{Belanger:2013oya}.

\section{Results}
\label{result}
In this section, we perform detail scanning of the parameter space of each of the generic model described in section \ref{pmssmmodels} subjected to three constraints - ATLAS eweakino search data in the $3l + \met$ channel at the LHC Run II, the observed DM relic density of the universe and the experimentally measured anomalous magnetic moment of the muon and identify the APS for each of them. We then discuss the prospects of discovery for these models through various multilepton channels for an integrated luminosity of 3000 fb$^{-1}$. We specifically emphasize on the $nl + \met$ channel with $n > 3$ that arises predominantly from the non-decoupled heavier eweakinos. 

We begin by estimating the SM backgrounds to all multilepton signals in subsection \ref{backgrnd}. Since the three compressed models introduced in section \ref{pmssmmodels} nicely highlight the importance of the heavier eweakinos, we first discuss the phenomenology of these models (see subsections \ref{comp}, \ref{modcomp} and \ref{hgsnlsp}). The following subsections deal with the remaining models.

\subsection{Estimation of the backgrounds to the multilepton signals}
\label{backgrnd}
In this subsection we obtain rough estimations of the backgrounds to the multilepton signals. For the $3l + \met$ signal, we take background obtained by the ATLAS Run II experiment in this channel \cite{Aaboud:2018jiw} and scale the number of events for the higher luminosity (3000 fb$^{-1}$). For simulating the $3l + \met$ signal in a pMSSM, we also follow the procedure of \cite{Aaboud:2018jiw}. For other signals namely $4l$, $ss3os1l$ (three same sign and one opposite sign $l$) and $5l$, suitable cuts are devised to control the SM background in each case (see Table \ref{tab}). The dominant SM processes contributing to the  multilepton final states are $t\bar{t}Z$, $ZZ$ and $VVV$ with $V = W^{\pm},Z$. 
\begin{table}[H]
\begin{center}
\begin{tabular}{|c||c|}
\hline
\hline
Channel & Cuts \\
\hline
$4l + \met$ &$N_l = 4$, $m_{SFOS}\not\in (81.2,101.2)$ GeV, $\met > 80$ GeV, $n_{b-jet} = 0$ \\
 & \\
$ss3os1l + \met$ &$N_l = 4$ with $Q_l \neq 0$, $\met > 80$ GeV \\
 & \\
$5l + \met$ &$N_l = 5$, $\met > 80$ GeV \\
\hline
\hline
\end{tabular}
\end{center}
\caption{The different choices of cuts for each type of multilepton signal.}
\label{tab}
\end{table} 
In Table \ref{tab} $N_l, Q_l$ are total number of isolated leptons in the final state and their total electric charge respectively and $m_{SFOS}$ is the invariant mass of a pair of same flavour opposite sign (SFOS) lepton pair. The main background in case of $4l + \met$ channel comes from pair productions of $Z$ boson and hence the invariant mass cut around the $Z$-window turns out to be very useful for reducing the background events. As the lepton multiplicity in the final state increases, the backgrounds become weaker and can be adequately suppressed by fewer cuts. For example, for the $5l + \met$ signal a moderate cut of 80 GeV on $\met$ is sufficient to make the background negligible. 

The total effective cross-section (i.e. the cross-section after all cuts) of the SM backgrounds in the $3l + \met$ channel listed in Table \ref{tab} is 0.261 fb. For the $nl + \met$ channels with $n > 3$, the total SM backgrounds are negligible. The strength of each multilepton signal is illustrated by two observables $\sigma_{eff}$ and $N_{BSM}$ where $\sigma_{eff}$ is the effective cross-section in the respective channels after passing all cuts and $N_{BSM}$ is the corresponding number of surviving signal events. As a rough guideline we require $N_{BSM} =$ 5 for discovery if the background is negligible. For the $3l + \met$ channel, however, we quote the signal significance $S/\sqrt{B}$ where $S$ is the number of signal events and $B$ is the number of corresponding background events which is nonzero.

\subsection{Compressed Light Higgsino Heavy Slepton (CLHHS ($\wt{W}$)) Model}
\label{comp}

\begin{figure}[h]
\centering
\includegraphics[width=0.6\textwidth]{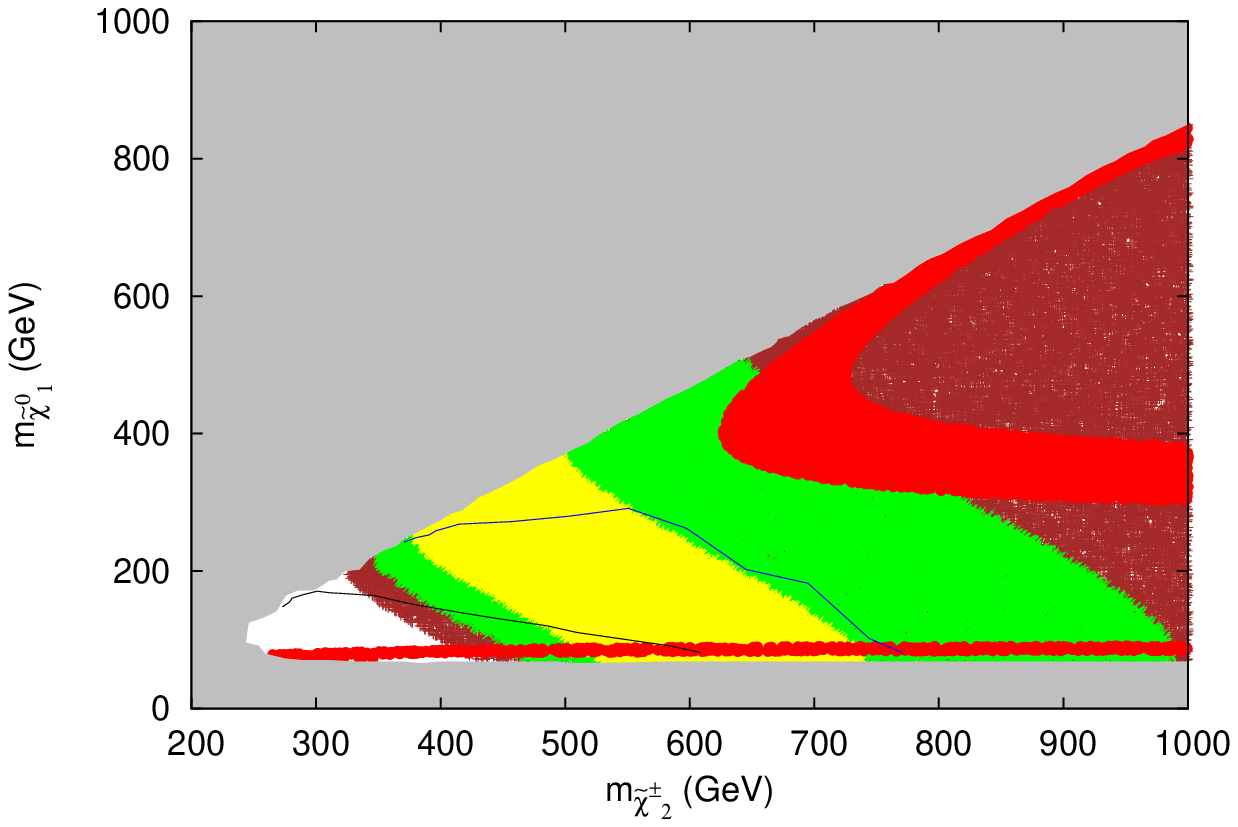}  
\caption{Exclusion contours in the $\mchtwopm-\mlspone$ plane in the Compressed Light Higgsino Heavy Slepton (CLHHS ($\wt{W}$)) model. The blue (black) line represents exclusion obtained by us (\cite{Chakraborti:2017vxz}) using the ATLAS $3l + \met$ search data from Run II (Run I). Colors and conventions are same as in Fig. \ref{fig}.}
\label{fig1}
\end{figure}

We first present the result of scanning the parameter space of the compressed model (section \ref{compmodel}, model (2.4 a)) by varying two gaugino masses in Fig. \ref{fig1}. Along the $x$-axis $\mchtwopm$ is varied while along the $y$-axis the variable is $\mlspone$ which is nearly degenerate with other lighter eweakino masses. The blue (black) contour represents the exclusion coming from $3l + \met$ data at 13 (8) TeV \cite{Aad:2014vma, Aaboud:2018jiw}. The Run II data rules out a larger part of the parameter space as compared to Run I data. For a LSP of mass around 80 GeV, the bound on $\mchtwopm$ is now extended upto $\approx 800$ GeV (previously it was nearly 600 GeV). Since we have illustrated the effect of compression by the choice $\mu = 1.05$ $M_1$, $\mlspone <$ 80 GeV is not allowed by the LEP lower bound on $\mchonepm$ \cite{lepsusy}. On the other hand, above $\mlspone \approx 350$ GeV (which was around 200 GeV for Run I), there is no bound on $\mchtwopm$. 
The eweakino search in the $3l + \met$ channel at Run II disfavours bulk of the bands allowed by the $a_{\mu}$ constraint at 1$\sigma$ and 2$\sigma$ levels for low $\mchtwopm$. But almost the entire 2$\sigma$ band in the high $\mchtwopm$ region survives. Although the red parabolic region allowed by the measured DM relic density remains unaffected by the Run II data, a large part of lower branch is excluded by the same. 
\begin{table}[H]
\begin{center}
\begin{tabular}{|c|c|c|c|c|c|c|c|c|c|c|c|}
\hline
\hline
\multicolumn{3}{|c|}{Mass} &Cross-section  &\multicolumn{2}{c|}{$3l + \met$} &\multicolumn{2}{c|}{$4l + \met$} &\multicolumn{2}{c|}{$ss3os1l + \met$} &\multicolumn{2}{c|}{$5l + \met$}\\
\cline{1-3} \cline{5-6} \cline{7-8} \cline{9-10} \cline{11-12}
$\mlspone$ &$\mchonepm$ &$\mchtwopm$ &in fb &$\sigma_{eff}^{3l}$ &$(S/\sqrt{B})_{3l}$ &$\sigma_{eff}^{4l}$ &$N_{4l}$ &$\sigma_{eff}^{ss3os1l}$ &$N_{ss3os1l}$ &$\sigma_{eff}^{5l}$ &$N_{5l}$ \\
\cline{1-12}
249.7 &290.2 &649.9 &156.7 &0.0423 &13.1 &0.0752 &225.6 &0.0282 &84.6 &0.0157 &47.0 \\
399.9 &440.7 &650.2 &40.14 &0.0064 &1.9 &0.0313 &93.9 &0.0144 &43.3 &0.0068 &20.5 \\
499.8 &527.9 &650.3 &23.64 &0.0054 &1.2 &0.0147 &43.9 &0.0083 &24.8 &0.0033 &9.9 \\
\hline
\hline
199.9 &239.8 &749.8 &298.0 &0.0805 &24.9 &0.0387 &116.2 &0.0059 &17.9 &0.0057 &17.1 \\
400.7 &445.4 &750.3 &33.42 &0.0144 &4.4 &0.0214 &64.2 &0.0070 &21.1 &0.0053 &16.1 \\
550.8 &591.0 &750.3 &13.59 &0.0031 &0.69 &0.0098 &29.4 &0.0043 &13.0 &0.0024 &7.3 \\
\hline
\hline
300.6 &344.4 &850.2 &80.86 &0.0307 &9.5 &0.0145 &43.7 &0.004 &12.1 &0.0032 &9.7 \\
400.6 &447.1 &849.9 &30.68 &0.0129 &3.9 &0.0117 &34.9 &0.0043 &12.9 &0.0012 &3.7 \\
500.2 &548.3 &850.0 &14.40 &0.0056 &1.7 &0.0095 &28.5 &0.0029 &8.6 &0.0017 &5.2 \\
\hline
\hline
350.7 &376.2 &500.3 &88.62 &0.0195 &3.8 &0.0691 &207.4 &0.031 &93.1 &0.0071 &21.3 \\
350.9 &393.9 &700.4 &53.56 &0.0198 &6.1 &0.0348 &104.4 &0.0134 &40.2 &0.0075 &22.5 \\
350.4 &396.0 &899.9 &47.59 &0.0186 &5.7 &0.0081 &24.3 &0.0038 &11.4 &0.0019 &5.7 \\
\hline
\hline
\end{tabular}
\end{center}
\caption{The masses and production cross-sections of all possible eweakino pairs for different BPs in the CLHHS ($\wt{W}$) model are given. For the trilepton signal in each case we display the significance ($S/\sqrt{B}$). The corresponding $\sigma_{eff}$ and total number of signal events (with negligible backgrounds) for each type of multilepton signal with $n>3$ are also shown. Masses and cross-sections are in GeV and fb respectively.}
\label{tab1}
\end{table} 

In Table \ref{tab1}, we showcase the results of our simulations of multilepton signals at $\sqrt{S} = 13$ TeV for an integrated luminosity of 3000 fb$^{-1}$ using BPs chosen from the APS. For clarity we have studied four groups of BPs all belonging to the APS shown in Fig. \ref{fig1}. For the first three groups, $\mchtwopm$ is fixed at 650, 750 and 850 GeV respectively while $\mlspone$ is varied. It is important to note that for $\mlspone >$ 350 GeV, the trilepton signal is below the observable level ($S/\sqrt{B} < 5$) irrespective of $\mchtwopm$. In most of such cases  one of the multilepton signals with $n > 3$ is likely to be the discovery channel ($N_{BSM} > 5$). On the other hand the last group of BPs illustrates that the $3l + \met$ signal improves for $\mlspone =$ 350 GeV even for $\mchtwopm$ as high as 900 GeV. Similar features have been observed for the moderately compressed model (see the next subsection).

\subsection{Moderately Compressed Light Higgsino Heavy Slepton (MCLHHS ($\wt{W}$)) Model}  
\label{modcomp}

\begin{figure}[h]
\centering
\includegraphics[width=0.6\textwidth]{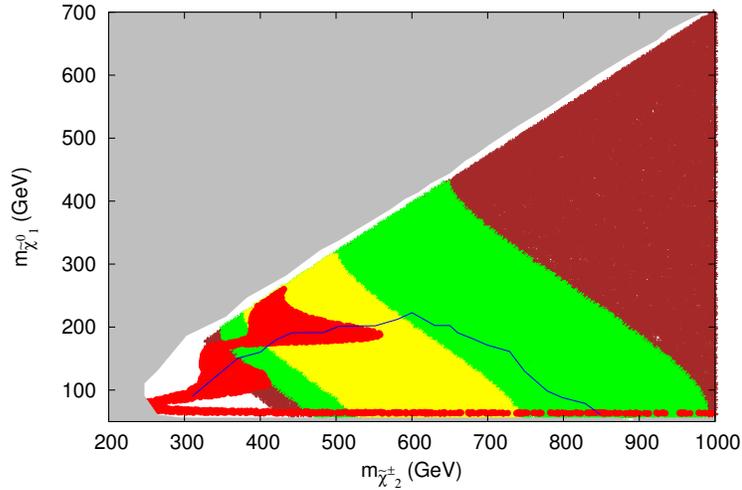}
\caption{Exclusion contour in the $\mchtwopm-\mlspone$ plane in the Moderately Compressed Light Higgsino Heavy Slepton (MCLHHS ($\wt{W}$)) model. The blue  line represents the exclusion obtained by us using the ATLAS $3l + \met$ search data from Run II. Colors and conventions are same as in Fig. \ref{fig}.}
\label{fig2}
\end{figure}

Fig. \ref{fig2} represents the result of scanning in the $\mchtwopm - \mlspone$ plane in the model with moderate compression (model (2.4 b)) illustrated by the choice $\mu = 1.3$ $M_1$. The blue line is the exclusion contour coming from the ATLAS $3l + \met$ data at Run II. For a LSP with mass around 60 GeV, $\mchtwopm$ below 850 GeV is disfavoured by the LHC search. LSP mass cannot be lowered further due to the lower bound on $\mchonepm$ coming from the LEP \cite{lepsusy} data. On the other hand above $\mlspone \approx$ 230 GeV, all $\chtwopm$ masses are allowed.

We pointed out in \cite{Chakraborti:2017vxz} that by relaxing the compression between $\mu$ and $M_1$, it is possible to get the DM relic density satisfying parameter space for lower values of $\mchtwopm$ which looks interesting from the perspective of sparticle searches. The red points in Fig. \ref{fig2} is consistent with the observed DM relic density of the universe. A significant fraction of the upper red patches lying in the low $\mchtwopm$ region is indeed allowed by the Run II data. However for the lower red band only the part with $\mchtwopm \gsim 850$ GeV is consistent with present LHC limits. For $\mchtwopm$ lying approximately in the range 500 - 600 GeV a significant part of the APS is consistent with both DM relic density and $a_{\mu}$ data (at 1$\sigma$ level). Also the 2$\sigma$ band of $a_{\mu}$ corresponding to larger values of $\mchtwopm$ is compatible with the $3l + \met$ data.

\begin{table}[H]
\begin{center}
\begin{tabular}{|c|c|c|c|c|c|c|c|c|c|c|c|}
\hline
\hline
\multicolumn{3}{|c|}{Mass} &Cross-section  &\multicolumn{2}{c|}{$3l + \met$} &\multicolumn{2}{c|}{$4l + \met$} &\multicolumn{2}{c|}{$ss3os1l + \met$} &\multicolumn{2}{c|}{$5l + \met$}\\
\cline{1-3} \cline{5-6} \cline{7-8} \cline{9-10} \cline{11-12}
$\mlspone$ &$\mchonepm$ &$\mchtwopm$ &in fb &$\sigma_{eff}^{3l}$ &$(S/\sqrt{B})_{3l}$ &$\sigma_{eff}^{4l}$ &$N_{4l}$ &$\sigma_{eff}^{ss3os1l}$ &$N_{ss3os1l}$ &$\sigma_{eff}^{5l}$ &$N_{5l}$ \\
\cline{1-12}
240.4 &325.4 &600.0 &137.9 &0.0469 &14.5 &0.0579 &173.7 &0.0152 &45.5 &0.0096 &28.9 \\
300.5 &397.3 &600.8 &74.0 &0.0244 &6.8 &0.0215 &64.4 &0.0126 &37.7 &0.0148 &44.4 \\
360.2 &461.9 &600.8 &48.98 &0.0279 &3.9 &0.01812 &54.4 &0.01029 &30.8 &0.0142 &42.6 \\
390.3 &484.7 &601.2 &44.03 &0.0136 &2.7 &0.0172 &51.5 &0.007 &21.1 &0.0075 &22.4 \\
\hline
\hline
110.4 &166.9 &800.4 &1249.0 &0.1124 &34.6 &0.0249 &74.9 &0.0125 &37.5 &0.0125 &37.5 \\
199.9 &278.7 &800.4 &218.3 &0.0458 &14.1 &0.0306 &91.7 &0.0109 &32.7 &0.0022 &6.5 \\
301.3 &406.1 &801.4 &54.88 &0.0324 &9.9 &0.0082 &24.7 &0.0021 &6.3 &0.0033 &9.9 \\
400.7 &529.7 &800.6 &20.76 &0.0079 &2.4 &0.0081 &24.3 &0.0048 &14.3 &0.0029 &8.7 \\
501.5 &646.7 &800.1 &10.7 &0.0031 &0.86 &0.0029 &8.9 &0.0019 &5.8 &0.0013 &3.8 \\
\hline
\hline
300.3 &371.8 &501.1 &115.3 &0.0265 &5.6 &0.0507 &152.2 &0.0165 &49.6 &0.012133 &36.4 \\
300.1 &401.1 &700.2 &58.15 &0.0221 &6.5 &0.0105 &31.4 &0.0077 &23.2 &0.0077 &23.2 \\
300.6 &406.5 &900.1 &53.21 &0.0176 &5.4 &0.0043 &12.8 &- &- &0.0016 &4.8 \\
\hline
\hline
\end{tabular}
\end{center}
\caption{The masses and production cross-sections of all possible eweakino pairs for different BPs in the MCLHHS ($\wt{W}$) model are given. For the trilepton signal in each case we display the significance ($S/\sqrt{B}$). The corresponding $\sigma_{eff}$ and total number of signal events (with negligible backgrounds) for each type of multilepton signal with $n>3$ are also shown. Masses and cross-sections are in GeV and fb respectively.}
\label{tab2}
\end{table}

Table \ref{tab2} shows the status of multilepton signals for different representative BPs. The BPs are bunched into three groups for reasons discussed in the last subsection. For a $\chtwopm$ with masses 600 and 800 GeV, the entire range of LSP masses considered gives potential $nl + \met$ signals with $n > 3$ for 3000 fb$^{-1}$ of integrated luminosity. But in most of the cases the trilepton signal is weaker compared to the other multilepton channels. On the other hand, keeping LSP mass fixed around 300 GeV, we have also varied $\mchtwopm$ and found that it is possible to get significantly large $3l + \met$ signal even for 900 GeV. Although for $\mchtwopm =$ 900 GeV, the $ss3os1l$ and $5l$ signals are already rather weak. 

\subsection{CLHHS ({$\wt{B} - \wt{W}$}) Model}
\label{hgsnlsp}

\begin{figure}[h]
\centering
\includegraphics[width=0.6\textwidth]{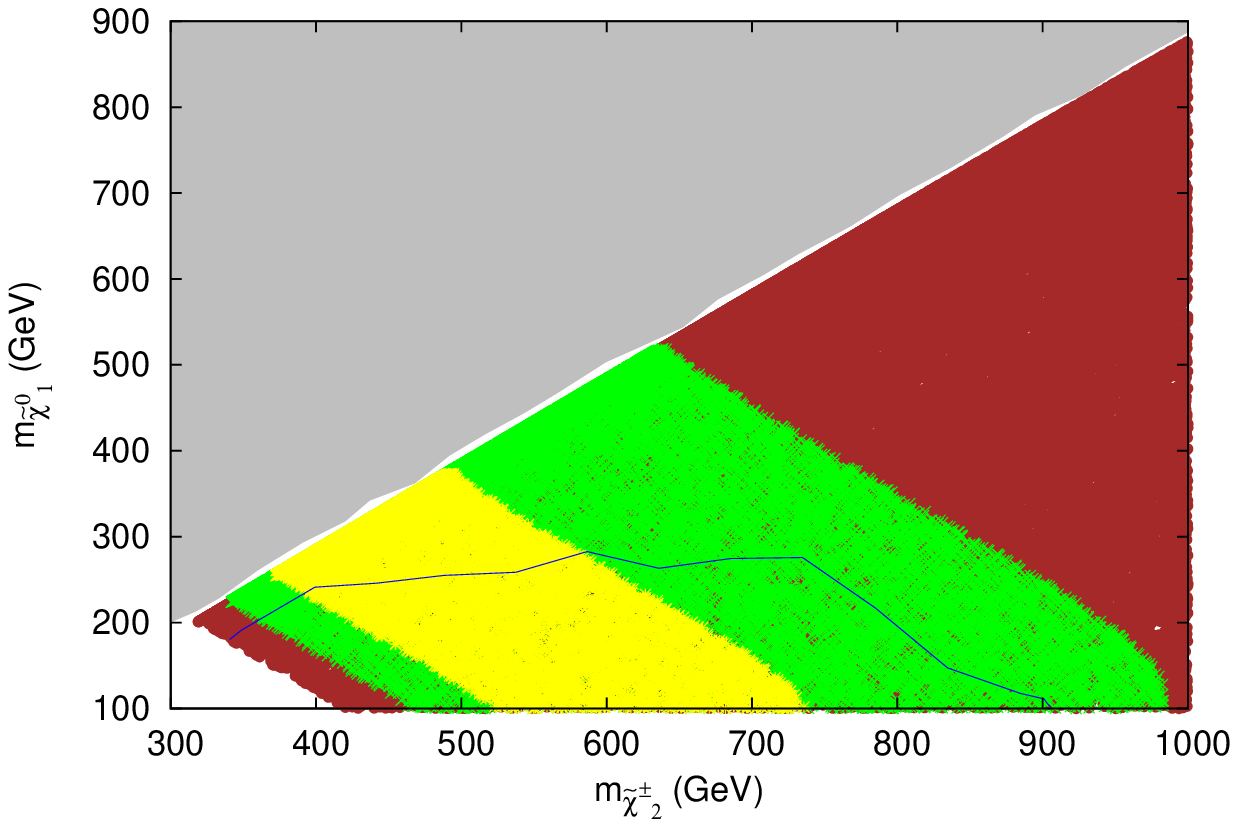}
\caption{Exclusion contour in the $\mchtwopm-\mlspone$ plane in the CLHHS ({$\wt{B} - \wt{W}$}) model. The blue line represents exclusion obtained by us using the ATLAS $3l + \met$ search data from Run II. Colors and conventions are same as in Fig. \ref{fig}.}
\label{fig3}
\end{figure}

The result of scanning the parameter space of this model (model (2.4 c)) is displayed in Fig. \ref{fig3}. The blue line is the exclusion contour obtained using the ATLAS $3l + \met$ search at Run II. The mass of $\chtwopm$ all the way upto 900 GeV is ruled out by the LHC data for a LSP with mass around 105 GeV. Note that, here LSP mass has a lower bound at around 105 GeV coming from the LEP data. This is because in this model the entire lighter eweakino spectrum is degenerate with mass controlled by higgsino parameter $\mu$ and hence the LEP bound on $\mchonepm$ is tantamount to a bound on $\mlspone$ (see subsection \ref{compmodel} model (2.4 c)). On the other hand, above $\mlspone \approx$ 290 GeV, there is no bound on $\mchtwopm$.

We also show the 1$\sigma$ (yellow) and 2$\sigma$ (green) allowed $a_{\mu}$ bands in the same plot. A fairly large part of these $a_{\mu}$ bands in the APS covering a wide range of $\mchtwopm$ is consistent with the present LHC Limit. Another point worth noting is that the APS as shown in Fig. \ref{fig3} does not contain any region consistent with the DM relic density data. This, however, is not surprising. It is well known that for a higgsino dominated LSP, DM relic density in the right ballpark value can be obtained only for high values of $\mlspone$ (e.g., $\mlspone$ around 1 TeV) \cite{Chattopadhyay:2003xi, Akula:2011jx} assuming single component DM. Our result agrees with this. In ref. \cite{Chakraborti:2017dpu}, authors have shown that this upper limit for higgsino DM mass can be relaxed if a small amount of slepton co-annihilation is present. Note that in  our model such a co-annihilation cannot occur as sleptons are much heavier than the LSP.
\begin{table}[H]
\begin{center}
\begin{tabular}{|c|c|c|c|c|c|c|c|c|c|c|c|}
\hline
\hline
\multicolumn{3}{|c|}{Mass} &Cross-section  &\multicolumn{2}{c|}{$3l + \met$} &\multicolumn{2}{c|}{$4l + \met$} &\multicolumn{2}{c|}{$ss3os1l + \met$} &\multicolumn{2}{c|}{$5l + \met$}\\
\cline{1-3} \cline{5-6} \cline{7-8} \cline{9-10} \cline{11-12}
$\mlspone$ &$\mchonepm$ &$\mchtwopm$ &in fb &$\sigma_{eff}^{3l}$ &$(S/\sqrt{B})_{3l}$ &$\sigma_{eff}^{4l}$ &$N_{4l}$ &$\sigma_{eff}^{ss3os1l}$ &$N_{ss3os1l}$ &$\sigma_{eff}^{5l}$ &$N_{5l}$ \\
\cline{1-12}
310.1 &316.9 &600.5 &140.31 &0.0449 &13.8 &0.0463 &138.9 &0.0014 &4.2 &0.0014 &4.2 \\
370.0 &377.4 &600.1 &80.21 &0.0104 &3.2 &0.0216 &64.9 &0.0016 &4.8 &- &- \\
\hline
\hline
330.2 &335.6 &700.6 &104.88 &0.0367 &11.3 &0.0231 &69.2 &- &- &- &- \\
380.2 &386.3 &700.1 &63.55 &0.0229 &7.1 &0.0178 &53.4 &0.0013 &3.8 &- &- \\
430.5 &437.2 &700.3 &41.55 &0.0108 &3.3 &0.0095 &28.6 &0.0017 &4.9 &- &- \\
\hline
\hline
280.8 &289.3 &500.1 &222.58 &0.0356 &10.9 &0.0935 &280.4 &0.0044 &13.4 &- &- \\
280.1 &285.4 &700.5 &190.24 &0.0609 &18.7 &0.0418 &125.6 &0.0019 &5.7 &- &- \\
280.2 &284.8 &800.5 &186.62 &0.0504 &15.5 &0.0186 &55.9 &- &- &- &- \\
\hline 
\hline
\end{tabular}
\end{center}
\caption{The masses and production cross-sections of all possible eweakino pairs for different BPs in the CLHHS ({$\wt{B} - \wt{W}$}) model are given. For the trilepton signal in each case we display the significance ($S/\sqrt{B}$). The corresponding $\sigma_{eff}$ and total number of signal events (with negligible backgrounds) for each type of multilepton signal with $n>3$ are also shown. Masses and cross-sections are in GeV and fb respectively.}
\label{tab3}
\end{table}  

In Table \ref{tab3} we present the results of multilepton signals for an integrated luminosity of 3000 fb$^{-1}$. Our investigation reveals that for $\mchtwopm \approx$ 700 GeV, the entire range of $\mlspone$ in the APS can be probed via $3l + \met$ and $4l + \met$ channel. Again, for a LSP of mass around 280 GeV, $\chtwopm$ as heavy as 800 GeV can lead to observable $3l/4l + \met$ signal. However note that, $ss3os1l + \met$ and $5l + \met$ channels produce weaker signals in most of the cases.

\subsection{Light Higgsino and Heavy Slepton (LHHS) Model} 
\label{lhhs}

\begin{figure}[h]
\centering
\includegraphics[width=0.6\textwidth]{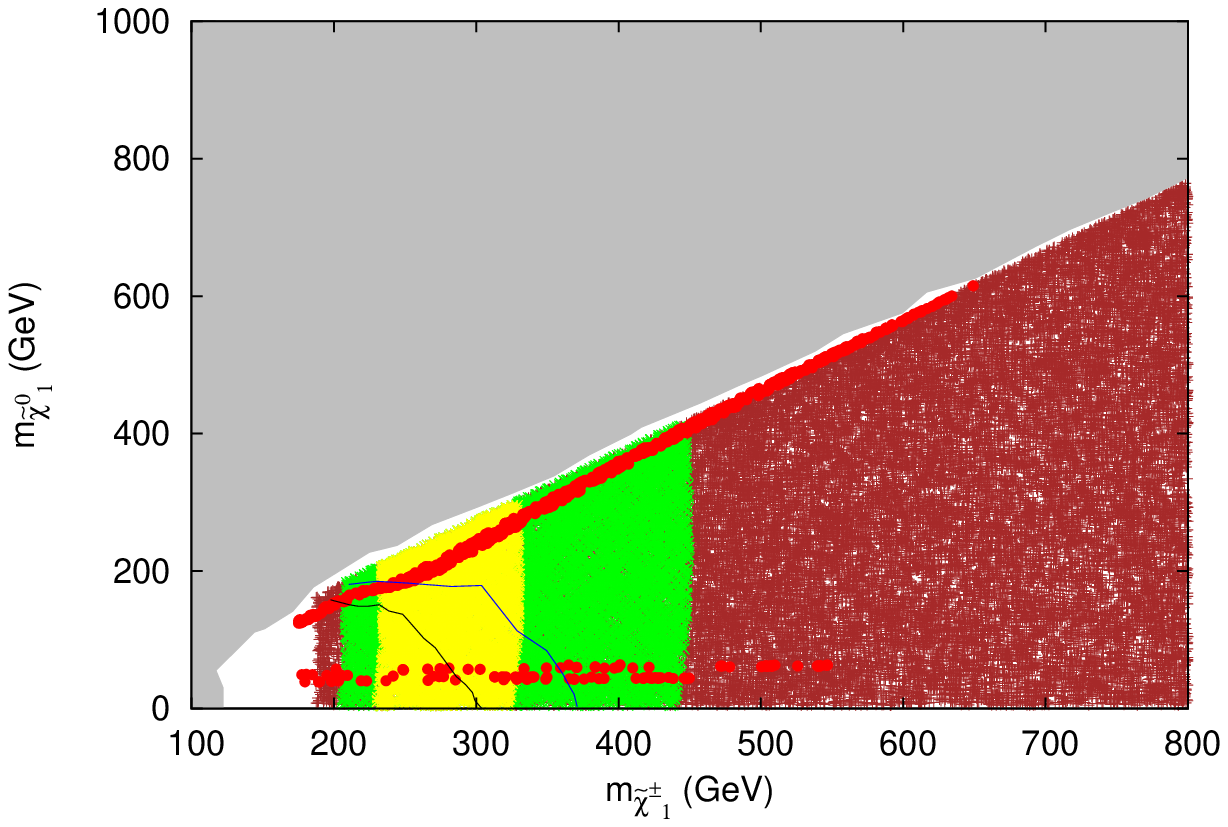}
\caption{Exclusion contours in the $\mchonepm-\mlspone$ plane in the Light Higgsino and Heavy Slepton (LHHS) model. The blue (black) line represents exclusion obtained by us (\cite{Chakraborti:2017vxz}) using the ATLAS $3l + \met$ search data from Run II (Run I). Colors and conventions are same as in Fig. \ref{fig}.}
\label{fig4}
\end{figure}

We delineate the APS of LHHS model (subsection \ref{hgsnmodel}, model (2.2 b)) in the $\mchonepm - \mlspone$ plane in Fig. \ref{fig4}. Run II data puts stronger constraint (the blue curve) on the APS compared to the Run I data (the black curve) \cite{Chakraborti:2017vxz}. For massless LSP, a $\chonepm$ with mass above $\gsim$ 370 GeV (the corresponding value of $\mchtwopm$ is $\gsim$ 600 GeV) is allowed whereas above $\mlspone \approx$ 200 GeV there is no bound on $\mchonepm$. Run II data eliminates a larger part of the lower DM band originating from $h$ and $Z$ resonances as compared to the Run I data. However almost the entire upper DM band survives except a tiny part. For $\mchonepm$ lying in the range 250 - 450 GeV, a large part of the APS is in agreement with both DM relic density and $a_{\mu}$ data (both at 1$\sigma$ and 2$\sigma$ levels).

We exhibit the results of multilepton searches for LHHS model in Table \ref{tab4}. The mode of presentation is the same as in earlier tables. For a $\chonepm$ with mass e.g. say 500 GeV, the entire range of LSP mass 50 - 450 GeV (see Fig. \ref{fig4}) allowed by the $3l + \met$ data can be probed at the LHC with $\mathcal{L} =$ 3000 fb$^{-1}$. On the other hand for a LSP of mass 300 GeV, good signal strength can be expected for almost each type of multilepton signal for $\mchonepm$ nearly upto 550 GeV (which corresponds to $\mchtwopm \approx$ 900 GeV). In some cases the $3l + \met$ signal again turns out to be weaker as compared to channels with higher lepton multiplicities. For higher values of $\mchonepm$, multilepton signal especially $ss3os1l$ and $5l$ signals get weaken rapidly. This can be understood easily as follows. In LHHS model, as sleptons masses are put between $\mchonepm$ and $\mchtwopm$ (see subsection \ref{hgsnmodel}, model (2.2 b)), only $\chtwopm (\lspfour)$ has direct decays into sleptons while the leptons can come from $\chonepm (\lsptwo,\lspthree)$ decays via SM gauge bosons with low BR. Therefore, the heavy EW sector is the main source of multileptons in this case. It was shown in \cite{Chakraborti:2017vxz} explicitly. Now, $\mchtwopm$ increases with increasing value of $\mchonepm$ and that in turn decreases the cross-section of heavy sector. As a result, one starts getting poor signals.

\begin{table}[H]
\begin{center}
\begin{tabular}{|c|c|c|c|c|c|c|c|c|c|c|c|}
\hline
\hline
\multicolumn{3}{|c|}{Mass} &Cross-section  &\multicolumn{2}{c|}{$3l + \met$} &\multicolumn{2}{c|}{$4l + \met$} &\multicolumn{2}{c|}{$ss3os1l + \met$} &\multicolumn{2}{c|}{$5l + \met$}\\
\cline{1-3} \cline{5-6} \cline{7-8} \cline{9-10} \cline{11-12}
$\mlspone$ &$\mchonepm$ &$\mchtwopm$ &in fb &$\sigma_{eff}^{3l}$ &$(S/\sqrt{B})_{3l}$ &$\sigma_{eff}^{4l}$ &$N_{4l}$ &$\sigma_{eff}^{ss3os1l}$ &$N_{ss3os1l}$ &$\sigma_{eff}^{5l}$ &$N_{5l}$ \\
\cline{1-12}
40.7 &380.7 &620.3 &86.86 &0.0669 &20.6 &0.0269 &80.8 &0.0165 &49.5 &0.0043 &13.0 \\
159.7 &380.7 &620.3 &86.16 &0.0534 &16.4 &0.0293 &87.8 &0.012 &36.2 &0.0103 &31.0 \\
321.3 &380.7 &620.3 &74.43 &0.0231 &7.1 &0.0372 &111.6 &0.0067 &20.1 &0.0082 &24.6 \\
\hline
\hline
49.98 &500.8 &795.9 &26.57 &0.0268 &8.3 &0.0074 &22.3 &0.0039 &11.9 &0.0027 &7.9 \\
199.7 &500.8 &795.9 &26.48 &0.0244 &7.5 &0.0087 &26.2 &0.0045 &13.5 &0.0026 &7.8 \\
400.3 &500.8 &795.9 &25.64 &0.0128 &3.8 &0.0053 &15.8 &0.0015 &4.5 &0.0035 &10.5 \\
\hline
\hline
300.4 &350.7 &576.9 &99.23 &0.0288 &7.9 &0.0635 &190.5 &0.0228 &68.5 &0.0039 &11.9 \\
300.3 &449.6 &720.8 &41.73 &0.0196 &6.1 &0.0108 &32.5 &0.0033 &10.0 &0.0037 &11.3 \\
300.4 &550.7 &869.6 &17.07 &0.0145 &4.4 &0.0051 &15.4 &0.0027 &8.2 &0.0005 &1.5 \\
\hline 
\hline

\end{tabular}
\end{center}
\caption{The masses and production cross-sections of all possible eweakino pairs for different BPs in the LHHS model are given. For the trilepton signal in each case we display the significance ($S/\sqrt{B}$). The corresponding $\sigma_{eff}$ and total number of signal events (with negligible backgrounds) for each type of multilepton signal with $n>3$ are also shown. Masses and cross-sections are in GeV and fb respectively.}
\label{tab4}
\end{table}  

\subsection{Light Higgsino and Light Left Slepton (LHLS) Model}  
\label{lhls}

\begin{figure}[h]
\centering
\includegraphics[width=0.6\textwidth]{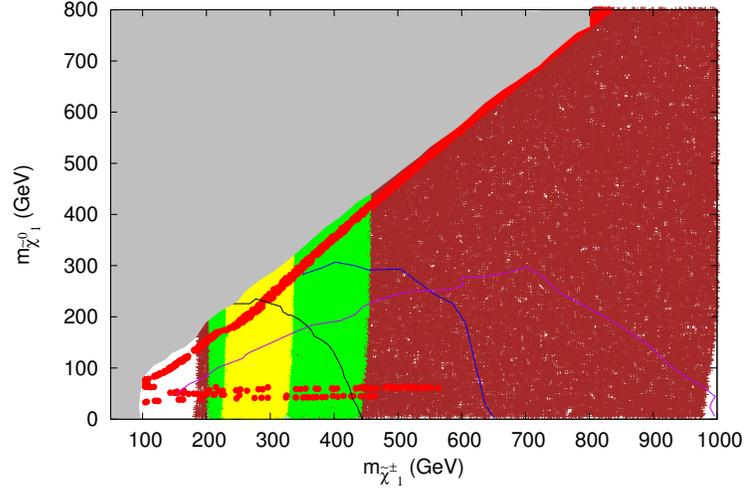}
\caption{Exclusion contours in the $\mchonepm-\mlspone$ plane in the Light Higgsino and Light Left Slepton (LHLS) model. The blue (black) line represents exclusion obtained by us (\cite{Chakraborti:2017vxz}) using the ATLAS $3l + \met$ search data from Run II (Run I). The exclusion from the direct slepton search (the magenta curve) is also shown. Colors and conventions are same as in Fig. \ref{fig}.}
\label{fig5}
\end{figure}

We show the exclusion contour (the blue curve) in the $\mchonepm$ - $\mlspone$ plane obtained by scanning the parameter space of the LHLS Model (see section \ref{hgsnmodel}, model (2.2 a)) in Fig. \ref{fig5}. The choice of the L-slepton masses is as in section \ref{hgsnmodel}. The constraints are significantly stronger than the ones obtained from the Run I data (the black curve) \cite{Chakraborti:2017vxz}. For example, the lower bound on $\mchonepm$ for a LSP with negligible mass is now extended from 450 GeV (Run I) to 650 GeV. The corresponding lower bound on $\mchtwopm$ is $\approx$ 1.01 TeV. Above $\mlspone \simeq$ 300 GeV, there is no bound on $\mchonepm$. In addition one can also put correlated bounds on $\mchonepm$ and $\mlspone$ coming from ATLAS slepton search\footnote{The slepton mass in the LHLS model is related to $\mchonepm$ and $\mlspone$ through the assumption $m_{\tilde{l}_L} = \frac{\mchonepm + \mlspone}{2}$.} at Run II \cite{Aaboud:2018jiw} of the LHC (see the magenta curve in Fig. \ref{fig5}). It is interesting to note that the bound on $\mchonepm$ for a massless LSP as obtained from the slepton search is around 1 TeV which is much stronger than that coming from direct eweakino searches at Run II.

The exclusion using Run II data depletes the bands allowed by the $a_{\mu}$ data severely leaving only a small fraction of the green $2\sigma$ band within the APS. The lower branch of the red region allowed by the DM relic density constraint, a part of which was allowed by the LHC Run I eweakino searches, are now  excluded by the Run II data. This has implications for the compatibility of this model and the DM direct detection data taken at its face value (see below). A significant portion of the upper red branch is still allowed.

\begin{table}[H]
\begin{center}
\begin{tabular}{|c|c|c|c|c|c|c|c|c|c|c|c|}
\hline
\hline
\multicolumn{3}{|c|}{Mass} &Cross-section  &\multicolumn{2}{c|}{$3l + \met$} &\multicolumn{2}{c|}{$4l + \met$} &\multicolumn{2}{c|}{$ss3os1l + \met$} &\multicolumn{2}{c|}{$5l + \met$}\\
\cline{1-3} \cline{5-6} \cline{7-8} \cline{9-10} \cline{11-12}
$\mlspone$ &$\mchonepm$ &$\mchtwopm$ &in fb &$\sigma_{eff}^{3l}$ &$(S/\sqrt{B})_{3l}$ &$\sigma_{eff}^{4l}$ &$N_{4l}$ &$\sigma_{eff}^{ss3os1l}$ &$N_{ss3os1l}$ &$\sigma_{eff}^{5l}$ &$N_{5l}$ \\
\cline{1-12}
309.9 &480.2 &763.6 &31.42 &0.0638 &19.6 &0.0223 &66.9 &0.0047 &14.1 &0.0028 &8.5 \\
359.8 &480.2 &763.6 &30.76 &0.0489 &13.5 &0.0225 &67.4 &0.0065 &19.4 &0.0037 &11.1 \\
410.1 &480.3 &763.8 &28.62 &0.0641 &9.1 &0.0346 &103.9 &0.0077 &23.2 &0.0054 &16.3 \\
\hline
\hline
259.6 &600.4 &940.5 &11.41 &0.0604 &18.5 &0.0073 &21.9 &0.0008 &2.4 &0.0009 &2.7 \\
349.8 &600.4 &940.5 &11.29 &0.0387 &11.9 &0.0048 &14.6 &0.0009 &2.7 &0.0008 &2.0 \\
550.2 &600.5 &940.9 &9.252 &0.0149 &2.1 &0.0077 &23.0 &0.0018 &5.3 &0.0012 &3.6 \\
\hline
\hline
400.1 &450.6 &720.4 &34.67 &0.0368 &8.1 &0.0319 &95.7 &0.0055 &16.6 &0.0028 &8.3 \\
400.4 &549.6 &865.6 &16.82 &0.0225 &6.3 &0.0056 &16.7 &0.0012 &3.5 &0.0015 &4.5 \\
400.5 &650.1 &1014.0 &7.68 &0.0230 &7.1 &0.0039 &11.7 &0.0008 &2.5 &0.0002 &0.5 \\
\hline
\hline
  
\end{tabular}
\end{center}
\caption{The masses and production cross-sections of all possible eweakino pairs for different BPs in the LHLS model are given. For the trilepton signal in each case we display the significance ($S/\sqrt{B}$). The corresponding $\sigma_{eff}$ and total number of signal events (with negligible backgrounds) for each type of multilepton signal with $n>3$ are also shown. Masses and cross-sections are in GeV and fb respectively.}
\label{tab5}
\end{table} 

Various multilepton signals in this model for an integrated luminosity of 3000 fb$^{-1}$ are displayed in Table \ref{tab5}. The $3l +\met$ signal is observable for almost the full set of BPs considered here. It also follows that $N_{BSM}$ exceeds 5 for all signals with $n \geq 3$ for a relatively low $\mchonepm = 480$ GeV for several choices of the LSP mass. However as $\mchonepm$ increases to 600 GeV, the cross-section for heavy eweakino pair production decreases rapidly (for this value of $\mchonepm$, we have $\mchtwopm (\mlspfour) \approx$ 940 GeV). The $ss3os1l$ and $5l$ signals that mainly come from heavy eweakino productions become weaker. The same features are seen when we vary $\mchonepm$ keeping $\mlspone$ fixed at 400 GeV. For the entire range considered by us the $4l+\met$ signal, which is not very common in the corresponding model with decoupled heavier eweakinos, is observable.

\subsection{Light Mixed and Light Left Slepton (LMLS) Model} 
\label{lmls}

\begin{figure}[h]
\centering
\includegraphics[width=0.6\textwidth]{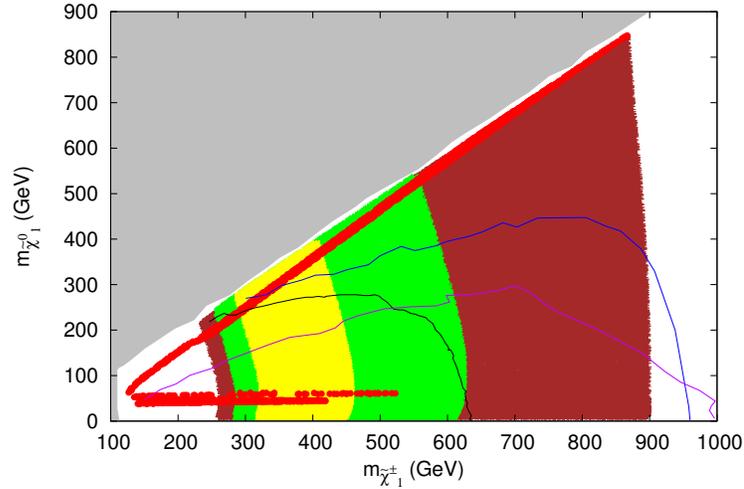}
\caption{Exclusion contours in the $\mchonepm-\mlspone$ plane in the Light Mixed and Light Left Slepton (LMLS) model. The blue (black) line represents exclusion obtained by us \cite{Chakraborti:2017vxz} using the ATLAS $3l + \met$ search data from Run II (Run I). The exclusion from the direct slepton search (the magenta curve) is also shown. Colors and conventions are same as in Fig. \ref{fig}.}
\label{fig6}
\end{figure}

The APS of the LMLS model in the $\mchonepm - \mlspone$ plane consistent with all constraints is shown in Fig. \ref{fig6}. The parameter space is tightly constrained in this case. The bound on $\mchonepm$ for a massless LSP coming from the Run II data (the blue curve) is $\approx$ 960 GeV. This limit on $\mchonepm$ differs from that in case of Run I (the black curve) \cite{Chakraborti:2017vxz} by atleast 300 GeV. On the other hand above $\mlspone = 450$ GeV, the LHC puts no constraint on the mass of $\chtwopm$. The magenta line represents the exclusion limit on $\mchonepm$ as a function of LSP mass coming from the LHC slepton search \cite{Aaboud:2018jiw}. The present LHC limits affect severely the part of the APS which is consistent with both DM relic density and $a_{\mu}$ data over a small region. A small part of 1$\sigma$ and 2$\sigma$ allowed $a_{\mu}$ bands lies beyond the Run II exclusion contour. The APS with $\mchonepm$ in the range 350 - 600 GeV is phenomenologically very interesting as it is allowed by both DM relic density and $a_{\mu}$ data. Although the upper DM band extends upto $\mchonepm \approx$ 900 GeV, the region with high $\mchonepm$ is likely to give poor multilepton signal at the high luminosity LHC (see below). Note that, the lower DM band was already ruled out by the Run I search. 

Table \ref{tab6} represents the result of multilepton signals in the LMLS model with the help of several BPs. For $\mchonepm = 550$ GeV, the entire allowed range of LSP masses (see Fig. \ref{fig6}) may be probed through multilepton channels with an integrated luminosity of 3000 fb$^{-1}$. In fact, even the channels like $ss3os1l + \met, 5l + \met$ which are unique features of non-decoupled heavy eweakinos yield large signals. For higher values of $\mchonepm$ (e.g. say 750 GeV), however, $ss3os1l$ and $5l$ signals are poor even with $\mathcal{L} =$ 3000 fb$^{-1}$. This again mainly happens due to large masses of the heavier eweakino sector that result into low cross-section. We get the same result when we vary $\mchonepm$ keeping LSP mass fixed at a particular value (say 500 GeV). The $4l + \met$ channel turns out to be the most promising for the rest of the BPs. 

\begin{table}[H]
\begin{center}
\begin{tabular}{|c|c|c|c|c|c|c|c|c|c|c|c|}
\hline
\hline
\multicolumn{3}{|c|}{Mass} &Cross-section  &\multicolumn{2}{c|}{$3l + \met$} &\multicolumn{2}{c|}{$4l + \met$} &\multicolumn{2}{c|}{$ss3os1l + \met$} &\multicolumn{2}{c|}{$5l + \met$}\\
\cline{1-3} \cline{5-6} \cline{7-8} \cline{9-10} \cline{11-12}
$\mlspone$ &$\mchonepm$ &$\mchtwopm$ &in fb &$\sigma_{eff}^{3l}$ &$(S/\sqrt{B})_{3l}$ &$\sigma_{eff}^{4l}$ &$N_{4l}$ &$\sigma_{eff}^{ss3os1l}$ &$N_{ss3os1l}$ &$\sigma_{eff}^{5l}$ &$N_{5l}$ \\
\cline{1-12}
400.0 &550.1 &664.3 &26.18 &0.0688 &19.1 &0.0259 &77.7 &0.0034 &10.2 &0.006 &18.1 \\
459.8 &550.4 &664.5 &25.76 &0.0551 &10.9 &0.0227 &68.0 &0.0059 &17.8 &0.0049 &14.7 \\
520.5 &550.7 &664.8 &22.32 &0.0112 &2.5 &0.0134 &40.2 &0.0024 &7.4 &0.0033 &10.0 \\
\hline
\hline
100.6 &750.9 &866.1 &5.761 &0.2898 &89.4 &0.0074 &22.1 &0.0005 &1.72 &0.0006 &1.9 \\
300.5 &750.3 &865.3 &5.809 &0.1922 &59.2 &0.0084 &25.1 &0.0006 &1.74 &0.0009 &2.6 \\
499.9 &750.0 &864.8 &5.813 &0.0532 &16.4 &0.007 &21.1 &0.0006 &1.92 &0.0011 &3.3 \\
699.8 &750.9 &865.5 &5.464 &0.0102 &2.2 &0.0074 &22.3 &0.0012 &3.6 &0.0015 &4.6 \\
\hline
\hline
500.0 &530.1 &644.5 &26.44 &0.0148 &3.3 &0.0148 &44.4 &0.0042 &12.7 &0.0032 &9.5 \\
499.9 &629.8 &744.2 &13.92 &0.0339 &6.7 &0.0128 &38.4 &0.0019 &5.8 &0.004 &12.1 \\
500.1 &829.9 &945.1 &3.342 &0.0177 &2.5 &0.0049 &14.8 &0.0005 &1.4 &0.0004 &1.3  \\
\hline
\hline  

\end{tabular}
\end{center}
\caption{The masses and production cross-sections of all possible eweakino pairs for different BPs in the LMLS model are given. For the trilepton signal in each case we display the significance ($S/\sqrt{B}$). The corresponding $\sigma_{eff}$ and total number of signal events (with negligible backgrounds) for each type of multilepton signal with $n>3$ are also shown. Masses and cross-sections are in GeV and fb respectively.}
\label{tab6}
\end{table} 

\section{Constraints from dark matter direct detection experiments}
\label{drctdtctn}

\begin{figure}[h]
\centering
\includegraphics[width=0.6\textwidth]{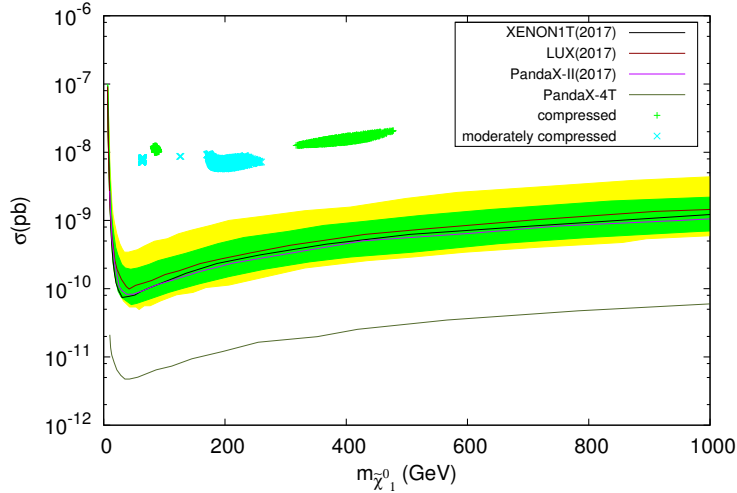}
\caption{Plot of spin-independent scattering cross-section $\sigma^{SI}$ for scattering of proton with $\lspone$ as a function of the LSP mass for the  compressed models ((2.4 a), (2.4 b) and (2.4 c)). Only the points which satisfy WMAP/PLANCK, $a_{\mu}$ upto the level of 2$\sigma$ and the LHC Run II constraints are used in the calculation. The exclusion contours for XENON1T, LUX, PandaX-II and PandaX-4T experiments are shown as black, red, magenta and green lines respectively. In green and yellow are shown 1$\sigma$ and 2$\sigma$ sensitivity bands respectively of the XENON1T data.}
\label{fig7}
\end{figure}

\begin{figure}[h]
\centering
\includegraphics[width=0.6\textwidth]{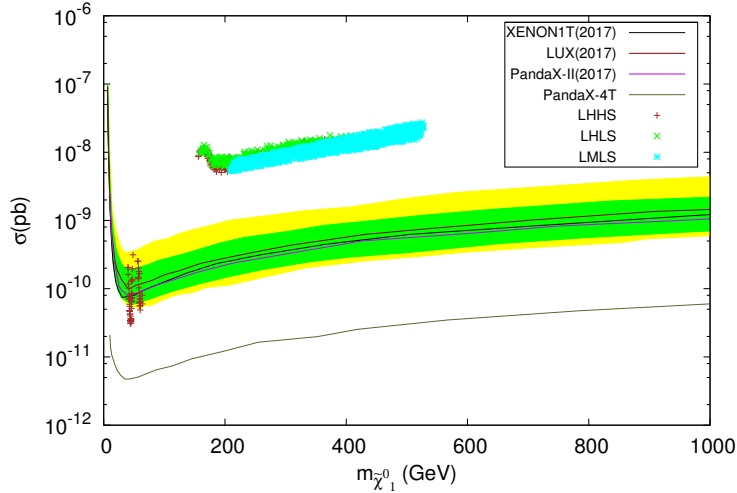}
\caption{Same as in Fig. \ref{fig7} but for LHHS, LHLS ans LMLS models.}
\label{fig8}
\end{figure}

In this section we study the models constrained in section \ref{result} in the light of the measured spin-independent DM nucleon scattering cross-section ($\sigma_{SI}$) by XENON1T \cite{Aprile:2017iyp}, LUX \cite{Akerib:2016vxi} and Panda \cite{Cui:2017nnn} experiments. However, in view of large uncertainties in the computation of $\sigma_{SI}$ due to theoretical as well as experimental inputs (see section \ref{methodology}), the relatively small differences between them are not very significant. From our scanning we take the points from the APS of each model, compute $\sigma_{SI}$ for them and compare the results with the upper bounds on $\sigma_{SI}$. Our results are shown in Figs. \ref{fig6}, \ref{fig7} and \ref{fig8}. In all figures the black curve represents the upper bound on $\sigma_{SI}$ as a function of the DM mass as obtained by the XENON1T experiment. The green and yellow regions represent 1$\sigma$ and 2$\sigma$ sensitivity bands respectively. The large widths of these bands reflect the statistical fluctuations in a typical low count experiment. The lowest curve shows the projected sensitivity of the PandaX-4T experiment \cite{Zhang:2018xdp}, which will be operational after the ongoing PandaX -II experiment.  

It follows from Fig. \ref{fig7} that both the compressed (CLHHS ($\tilde{W}$)) and the moderately compressed (MCLHHS ($\tilde{W}$)) models 
predict $\sigma_{SI}$ far above the experimental upper bounds. Thus these models can only survive provided the computed values of $\sigma_{SI}$  are overestimated by a large factor - a possibility that cannot be ruled out a priori. This can happen if e.g., the DM density in the neighbourhood of the earth, which has not been directly measured, turns out to be unexpectedly small. It may be recalled that only the average value  of this density over a astronomically large volume with the sun at the centre has been measured experimentally. Other uncertainties as discussed in subsection \ref{constraints} leave open the possibility that $\sigma_{SI}$ could be even further suppressed. Thus conclusions based on Fig. \ref{fig7} should not to be taken at their face values.

From Fig. \ref{fig8} it can be seen that LHLS and LMLS models are also disfavoured. However they cannot be ruled out with confidence thanks to the uncertainties in the computation of $\sigma_{SI}$ as discussed in the last paragraph. It is interesting to note that the LHHS model is still consistent with the DM direct detection data even if Fig. \ref{fig8} is taken at its face value. This happens in a part of the APS where the DM relic density is produced by the LSP pair annihilation into the Higgs boson. Similar parameter spaces in other models are now ruled out by the LHC Run II data.  
\begin{figure}[h]
\centering
\includegraphics[width=0.6\textwidth]{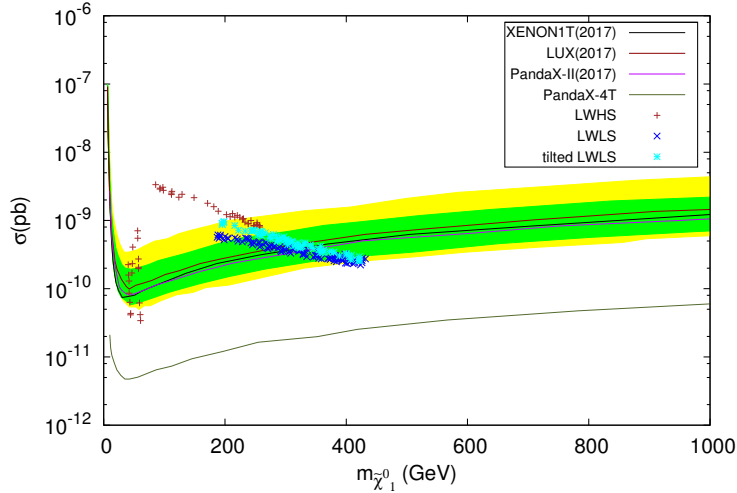}
\caption{Same as in Fig. \ref{fig7} but for various LW type models.}
\label{fig9}
\end{figure}  

We also note in passing that several LW models are also consistent with the direct detection data (see Fig. \ref{fig9}).

\section{Conclusions}
\label{conclusion}
In conclusion we reiterate that the search for the heavier eweakinos could be an important programme during the LHC after the current long shutdown. The searches for the hadronically quiet multilepton ($nl + \met$, $n > 3$ ) signals may even be the SUSY discovery channels if the lighter eweakinos have a compressed spectrum. 

In order to reach this conclusion we have carried out the following analyses. We first constrain the full eweakino sector of several generic pMSSMs, described in section \ref{pmssmmodels}, using the ATLAS model independent upper bound on the number of any BSM event from trilepton searches at Run II of the LHC \cite{Aaboud:2018jiw} (for a summary of models studied in this paper see subsection \ref{paramsummary} and Table \ref{summarytab}). We do not employ the often used ad hoc assumption that the heavier eweakinos are decoupled. As explained in section \ref{pmssmmodels} the phenomenology of the heavier eweakinos are particularly important in the light higgsino (LH) models (see subsection \ref{hgsnmodel}) where they ($\chtwopm$ and $\lspfour$ ) are dominantly winos. In this scenario  the lighter eweakinos ($\chonepm$, $\lsptwo$ and $\lspthree$) are higgsino dominated while the LSP is either a higgsino or bino-higgsino admixture. The exclusion contour obtained in a model also depends sensitively on the hierarchy between the slepton and eweakinos masses. Accordingly we have worked in two scenarios i) LHLS (model (2.2 a)) and ii) LHHS models (model (2.2 b)). In addition we have also considered the LMLS model (section \ref{mixedmodel}) where the lighter and heavier eweakinos - other than the LSP - are admixtures of wino and higgsino eigenstates. For the smallest allowed  LSP mass in each model the lower bounds on $\mchonepm$ are 650 GeV (Fig. \ref{fig5}), 370 GeV (Fig. \ref{fig4}) and 960 GeV (Fig. \ref{fig6}) respectively. All of them are significantly weaker than the ATLAS Run II limit of 1150 GeV (see Fig. \ref{fig}) for negligible LSP mass obtained in a simplified model similar to the LWLS model (model (2.1 a), subsection \ref{winomodel}) with decoupled heavier eweakinos. This indicates once more that the prospect of observing interesting physics involving relatively low mass eweakinos  in the LH models looks  brighter. The corresponding limits on $\mchtwopm$ are 1.01 TeV, 600 GeV and 1.07 TeV respectively. It also follows that the weakest exclusion from the LHC Run II data occurs in the LHHS model. As a result the APS, consistent with all constraints discussed in section \ref{constraints}, is quite large in this model. We also note in passing that the prediction of this model for $\sigma_{SI}$ is consistent with all DM direct detection data (section \ref{drctdtctn}).

It was emphasized in refs. \cite{Datta:2016ypd, Chakraborti:2017vxz} based on the ATLAS Run I data that the heavier eweakinos attain special significance if the lighter eweakino spectrum is compressed so that only weak signals involving mostly soft particles can emanate from them. Keeping this in view we have studied three compressed models i) CLHHS ($\wt{W}$) model (model (2.3 a), section \ref{compmodel}), ii) MCLHHS ($\wt{W}$) (model (2.3 b), section \ref{compmodel}) and iii) CLHHS ($\wt{W}- \wt{B}$) (model (2.3 c), section \ref{compmodel}). The exclusion contour and the APS for each model using the ATLAS Run II data are shown in Figs. 2, 3 and 4 respectively. For the lowest LSP mass allowed by the LEP data, the lower bounds on $\mchtwopm$ in these compressed models are 775 GeV, 850 GeV and 900 GeV respectively. On the other hand there is no constraints even from the Run II data for LSP masses above 200 - 300 GeV in any of these three compressed model.

The prospects of observing multilepton ($nl + \met, n=3,4,5$) signatures at the high luminosity LHC (3000 fb$^{-1}$) in different models are shown in Tables \ref{tab1} - \ref{tab6} using BPs. These points belong to the APS of the respective models constrained as described above.  As already noted, in the compressed models (see Tables \ref{tab1} - \ref{tab3}) the signals for n = 3 turn out to be  rather poor especially for relatively high LSP masses ( $> 350-400$ GeV). In such cases one of the search channels with $n > 3$ could be the discovery channel even for higher LSP masses. In particular the signal with n = 4 appears to be rather promising. Depending on the LSP mass, $\mchtwopm$ upto 1 TeV can be probed. For the non-compressed model all multilepton channels appear to be relevant provided the  LSP mass is around 400 - 450 GeV or smaller(see Tables \ref{tab4} - \ref{tab6}).

As discussed in section \ref{drctdtctn} the LHHS model deserves some attention since it's prediction for $\sigma_{SI}$, taken at its face value, is consistent with the upper bound on this cross-section measured by the DM direct detection experiments \cite{Akerib:2016vxi, Aprile:2017iyp, Cui:2017nnn} (Fig. \ref{fig7}). The predictions of all other LH type models violate the above bound by large factors (see Figs. \ref{fig7}, \ref{fig8}). We note in passing that the LW type models look better in this respect (see Fig. \ref{fig9}). Whether the computed $\sigma_{SI}$ should be taken at its face value is, however, not at all clear. This is because of several  inputs in the calculation which involve  large uncertainties (see subsection \ref{constraints} and references there in). We, therefore, refrain from spelling the final verdict based on the experimental upper bound on $\sigma_{SI}$.

{ \bf Acknowledgments :} The research of AD was supported by the Indian National Science Academy, New Delhi. NG acknowledges financial support from a DST SERB grant.


\end{document}